\documentclass[fleqn,twoside]{article}
\usepackage{espcrc2}

\newcommand{\AmS}{{\protect\the\textfont2
  A\kern-.1667em\lower.5ex\hbox{M}\kern-.125emS}}

\newcommand{\cc}{\cite}

\newcommand{\be}{\begin{equation}}
\newcommand{\ee}{\end{equation}}

\def\ex{\hbox{e}}

\def\<{\langle}
\def\>{\rangle}

\def\ch{\cosh}

\def\a{\alpha}
\def\b{\beta}
\def\g{\gamma}
\def\G{\Gamma}
\def\d{\delta}

\def\l{\lambda}
\def\s{\sigma}
\def\r{\rho}

\def\c{\chi}
\def\m{\mu}
\def\n{\nu}

\def\vf{\varphi}
\def\({\left(}
\def\[{\left[}
\def\){\right)}
\def\]{\right]}

\def\cos{\hbox{cos}}
\def\sin{\hbox{sin}}

\def\Tr{\hbox{Tr}}
\def\pa{\mathcal P}

\title{Study of instanton effects \\ in electromagnetic quark form factor at high energy}

\author{A.E. Dorokhov\address[MCSD]{Joint Institute for Nuclear Research\\
BLTP JINR, RU-141980 Dubna, Russia}%
        \thanks{Email: {\tt dorokhov@thsun1.jinr.ru}} ,
        I.O. Cherednikov\addressmark[MCSD]\address{Dipartimento di
Fisica, Universit\`a di Cagliari and INFN Sezione di Cagliari
\\ C.P. 170, I-09042  Monserrato (CA), Italy}\thanks{Email: {\tt igorch@thsun1.jinr.ru} }}

\begin{document}

\begin{abstract}
The detailed analysis of nonperturbative contributions to the
electromagnetic quark form factor is performed within the
framework of the instanton liquid model (ILM) of the QCD vacuum.
The method of the path-ordered Wilson exponentials is applied to
evaluate explicitly the instanton
corrections. By using the Gaussian interpolation of the
constrained instanton solution, it is shown that the instantons
yield  the logarithmic corrections to the amplitudes in high
energy limit which are exponentiated in small instanton density
parameter. \vspace{1pc}
\end{abstract}

\maketitle

\section{INTRODUCTION}

The necessity of a nonperturbative input for effective explicit
calculations of hadronic processes even in high energy domain is
an important reason for development of nonperturbative methods in
QCD \cc{pQCD}. The nonperturbative effects are naturally related
to the nontrivial structure of the QCD vacuum. In the last
decades, a great progress has been made in study of the QCD ground
state and a number of important results have been obtained that
connect the properties of the vacuum with the hadron
characteristics treating the QCD vacuum in the framework of the
instanton liquid model \cc{ILM,DP,DK,REV}. The idea that the
nontrivial vacuum structure can be relevant as well in high energy
hadronic processes was explicitly formulated in the context of the
so-called soft Pomeron problem in Refs. \cc{NL,LN}, and
further developed using the eikonal approximation and the method
of Wilson path-ordered exponentials in Refs. \cc{NACH,KRP}. Recently,
we investigated the instanton induced effects in high energy regime for the
electromagnetic (EM) quark form factor in the {\it weak-field approximation}, Ref. \cc{DCH1}.
In the present work, we report our results on evaluation of the
{\it total instanton contribution} to this quantity.

\section{EVOLUTION EQUATION FOR EM QUARK FORM FACTOR}

The electromagnetic Dirac quark form factors are determined via
the elastic scattering amplitude of a quark in an external EM
field: \be  \mathcal M_\m =F_q\[(p_1-p_2)^2\] \bar u(p_1) \g_\m
v(p_2) \ , \label{eq:ampl} \ee where $u(p_1), \ v(p_2)$ are the
spinors of outgoing and incoming quarks.
It is assumed that both the momentum transfer $- t = Q^2 = (p_1 -p_2)^2$ and the total
center-of-mass energy $s = (p_1 +p_2)^2$ are large compared to the quark mass, that is:
$ (p_1p_2) \gg p_{1,2}^2 =m^2 $, or $ \ch \c \gg
1 $ .

Using the classification of the diagrams with respect to the momenta
carried by their internal lines, the resummation of all
logarithmic terms coming from the soft gluon subprocesses allows
us to express the ``soft'' part of the form factor, $F_q$, in terms of the vacuum average of the gauge
invariant path-ordered Wilson exponential (within the eikonal approximation) \cc{ALLLOGA,KRON,MMP,Pol}
\be W (C_\c, \a_s)  =  \frac{1}{N_c} \Tr \Big< \pa \ex^{i g \int_{C_\c} \! d x_{\m} \hat A_{\m} (x)
} \Big\>_{0} \ .   \label{1a} \ee In Eq. (\ref{1a}), the
integration path corresponding to the considered process goes
along the closed contour $C_\c$: the angle (cusp) with infinite
sides. We parameterize the integration path $C_\c=\{z_\mu(t);
t=[-\infty,\infty]\}$ as follows \be z_{\mu}(t)=\left\{
\begin{array}
[c]{c}%
v_{1}t \ ,\qquad-\infty<t<0 \ ,  \\
v_{2}t \ ,\qquad0<t<\infty \ .
\end{array} \right. \label{path}\ee The Wilson integral (\ref{1a}) is
multiplicatively renormalizable \cc{KRC1,WREN}, therefore,
we can define the cusp anomalous dimension $\G_{cusp}$:
$ \G_{cusp} (\c; \a_s) = - \m \frac{d}{d \m} \ln W(\c; \m^2/\l^2, \a_s)$ ,
which determines the high-energy asymptotics of the form factor \cc{KRON}.
Here $\bar \m^2$ is the UV cutoff, $\m^2$ is the normalization point, and $\l^2$ is the IR cutoff.
In one-loop order of perturbative expansion, the cusp anomalous dimension is given by \cc{KRON}: $
{\G_{cusp}^{(1)}} ( \a_s) = \frac{\a_s}{\pi} C_F $ . In what
follows, we explicitly calculate the nonperturbative part $W_{np}$
within the ILM. The latter plays a role of initial conditions for perturbative evolution.

\section{INSTANTON INDUCED CORRECTIONS}

Let us consider the instanton contribution to the evolution equation for the form factor.
The instanton field has the
general form \be \hat A_\m (x; \r) =  \frac{1}{g}  {\hbox{\tt R}}^{ab} \s^a {\eta^\pm}^b_{\m\n} (x-z_0)_\n \vf
(x-z_0; \r) , \label{if1} \ee where $\varphi (x)$ is the gauge
dependent profile function, ${\hbox{\tt R}}^{ab}$ is the color
orientation matrix, $\s^a$'s are the Pauli matrices,
${\eta^\pm}^a_{\m\n}=\varepsilon_{4a\m\n}\mp(1/2)\varepsilon_{abc}\varepsilon_{bc\m\n}$
are 't Hooft symbols, and $(\pm)$ corresponds to the instanton or
antiinstanton solution. The averaging of the Wilson operator over
the nonperturbative vacuum is performed by the integration over
the coordinate of the instanton center $z_0$, the color
orientation and the instanton size $\r$.  The measure for the
averaging over the instanton ensemble reads $dI = d{\hbox{\tt R}}
\ d^4 z_0 \ dn_\r $, where $ d{\hbox{\tt R}}$ refers to the
averaging over color orientation, and $dn_\r$ depends on the
choice of the instanton size distribution. After averaging over
all possible embeddings of $SU_c(2)$ into $SU_c(3)$ \cite{SVZ80},
we write the Wilson integral (\ref{1a}) over the contour
(\ref{path}) in the single instanton approximation in the form:
$$ w_I(\c) =     \frac{1}{3}   \Tr \Big\< \frac{4}{9}\   \({\hbox{\tt I}} \times
{\hbox{\tt I}}\) \cos{\a_1
} \cos{\a_2}\   + \frac{1}{8} \(\l^A \times \l^A\) \cdot $$ \be
\[\frac{1}{3}\cos{\a_1} \cos{\a_2} - \hat n_1^a\hat n_2^a \ \sin{\a_1} \sin{\a_2}
 \]  \Big\>_0 ,
\ee
where $(i=1,2)$ \be \hat
n^a_i  = \frac{(-1)^i}{s(v_i,z_{0})} {\eta^\pm}^a_{\m\n} v_i^\mu
z_0^\nu \ , \label{iin} \ee
\be \a_i = s_i \cdot \int_0^\infty \! d\l \vf\[(
v_i\l +(-1)^iz_0)^2; \r_c \]  \ , \label{it2}
\ee
and $ s^2_i = z_0^2 -(v_i z_0)^2 $ ,  $(v_1v_2) = \cosh \c$ in Minkowski
geometry. After evaluating the traces, the resulting gauge
invariant contribution to the Wilson loop of the single instanton
reads \cc{DCH04}
$$
w_I(\chi) =   \frac{2}{3} \int dn_\r \[w_c^I(\c) + w_s^I(\c) - w_c^I(0)-w_s^I(0)\]
$$
\be w_c^I(\chi) = \int \! d^4 z_0 \ \cos \ \a_1  \cos \ \a_2 \   \ee
\be w_s^I(\chi) =  - \int \! d^4 z_0 \ (\hat n^a_1\hat n^a_2) \ \sin \ \a_1 \sin \ \a_2 \
\ee
where the normalized color correlation
factor is \be \hat n^a_1\hat n^a_2 =  -\frac{ \eta _{\m\n}^{a
}v_{1}^{\m}z_{0}^\n \eta _{\r \s}^{a}v_{2}^{\r
}z_{0}^\s} {s_1 s_2} \ . \ee

In realistic instanton vacuum model there are two essential
effects: stabilization of the instanton density with respect to
unbounded expansion of instantons in size and screening of
instantons by surrounding background fields. To take into account
these features we approximate first the narrow instanton size
distribution by the $\d$- function: $ dn_\r = n_c \d(\r-\r_c)
d\r $ , where the model parameters are \cc{REV}:
\be { n_c} \approx 1 fm^{-4}, \ \ { \r_c} \approx 1/3 fm
\label{param} \ , \ee and assume that the integration over the
instanton size is already performed. The screening effect modifies
the instanton shape at large distances leading to the constrained
instantons \cite{DEMM99}. To take into account this screening and
to have also simpler analytical form for $w_I(\chi)$, we use the
{\it Gaussian Anzatz }  for the instanton profile function \be \varphi
_{G}(x^{2})=\frac{1}{\rho_c^2}\ex^{-x^{2}/\rho_c ^{2}} \ .
\label{Inst_Profile_G} \ee The parameters in this expression are
fixed by the requirement of reproducing the vacuum average $\Big\<
g^2A_\mu^a(0)A_\mu^a(0)\Big\> = 12\pi^2\rho_c^2n_c \ . $

Performing tedious calculations (for technical details, see Ref.
\cc{DCH04}), we find that the total expression
for the quark form factor at large-$Q^2$ with the one-loop
perturbative contribution and the nonperturbative contribution
found in the instanton model reads: \be \frac{F_q \[Q^2\]}{F_q \[Q_0^2\]} =
\ex^{- \frac{2C_F}{\b_0} \ln Q^2 \ln \ln Q^2  - \ln Q^2
\(B_{inst}^{LOG} - \frac{2C_F}{\b_0} \)}
  \label{eq:final} \ee where the {\it all-order} instanton induced correction calculated
in the Gaussian approximation reads
\be
B_{inst}^{LOG} = - 1.0053 \ \frac{\pi^2  n_c{
\rho_c}^4}{12} \ . \label{bi}
\ee
Thus, the instanton induced effects (for the Gaussian simulation of instanton profile
function)  yield the logarithmic, {\it i.e.,} sub-leading,
correction to the high energy behavior of quark EM form factor,
with the numerical coefficient smaller then that of corresponding
perturbative term $ B_{inst}^{LOG} \ll B_{pert}^{LOG} = \frac{2C_F}{\beta_0} $ .
Comparing the results Eqs. (\ref{eq:final}) and (\ref{bi}) with the weak-field calculations
in the Ref. \cc{DCH1}, we conclude that the latter deliver the reasonable approximation, at least in the
case of Gaussian profile function.

\section*{ACKNOWLEDGEMENTS}
The authors thank the Organizers of Diffraction'04 for hospitality
and financial support. The work is partially supported by RFBR
(Grant nos. 04-02-16445, 03-02-17291, 02-02-16194), Russian
Federation President's Grant no. 1450-2003-2, and INTAS (Grant no.
00-00-366).

\end{document}